\documentclass{jpp}

\usepackage[utf8]{inputenc}
\usepackage{latexsym}
\usepackage{graphics}
\usepackage{graphicx}
\usepackage{subcaption}
\usepackage{amsmath}
\usepackage{amssymb}
\usepackage{color}
\usepackage[T1]{fontenc}
\usepackage[english]{babel}
\newcommand{\dime}{0.5\textwidth}
\newcommand{\ie}{\emph{i.e., }}
\newcommand{\omp}{\omega_p}
\newcommand{\reff}[1]{(\ref{#1})}
\newcommand{\eref}[1]{Eq.\reff{#1}}
\newcommand{\erefs}[1]{Eqs.\reff{#1}}
\newcommand{\figref}[1]{Fig.\ref{#1}}

\newcommand{\citer}[1]{\cite{#1}}
\newcommand{\citers}[1]{\cite{#1}}

\newcommand{\aes}{\mathrm{AE}}

\title{Contributions to the linear and non-linear theory of the beam-plasma interaction}

\author{N. Carlevaro\aff{1,2}, M. Del Prete\aff{3}, G. Montani\aff{1,3} \and
F. Squillaci\aff{3}}

\affiliation{
\aff{1}ENEA, Fusion and Nuclear Safety Department, C. R. Frascati,\\ Via E. Fermi 45, 00044 Frascati (Roma), Italy
\aff{2}Consorzio RFX, Corso Stati Uniti 4, 35127 Padova, Italy
\aff{3}Physics Department, ``Sapienza'' University of Rome, P.le Aldo Moro 5, 00185 Roma, Italy
\aff{4}INFN - Rome section, P.le Aldo Moro 2, 00185 Roma, Italy
}

\begin{document}
\maketitle

\begin{abstract}We focus our attention on some relevant aspects of the beam-plasma instability in order to refine some features of the linear and non-linear dynamics. After a re-analysis of the Poisson equation and of the assumption dealing with the background plasma in the form of a linear dielectric, we study the non-perturbative properties of the linear dispersion relation, showing the necessity for a better characterization of the mode growth rate in those flat regions of the distribution function where the Landau formula is no longer predictive. We then upgrade the original $N$-body approach in \citer{OWM71}, in order to include a return current in the background plasma. This correction term is responsible for smaller saturation levels and growth rates of the Langmuir modes, as result of the energy density transferred to the plasma via the return current. Finally, we include friction effects, as those due to the collective influence of all the plasma charges on the motion of the beam particles. The resulting force induces a progressive resonance detuning, because particles are losing energy and decreasing their velocity. This friction phenomenon gives rise to a deformation of the distribution function, associated with a significant growth of the less energetic particle population. The merit of this work is to show how a fine analysis of the beam-plasma instability outlines a number of subtleties about the linear, intermediate and late dynamics which can be of relevance when such a system is addressed as a paradigm to describe relevant nonlinear wave-particle phenomena \citep{ZCrmp}.
\end{abstract}


\section{Introduction}
A fundamental model of plasma physics consists of the beam-plasma instability, originally studied in \citers{OWM71,OM68,MK78,TMM94} (see also \citer{EEbook}), mainly having in mind real experiments, while it is currently a scenario to mimic some of the most relevant features of the interaction between fast ions and the Alfv\'en spectrum present in tokamak devices \citep{BB90a,BB90b,BB90c,BS11}.

The basic mechanism underlying the beam-plasma instability is the inverse Landau damping mechanism \citep{LP81,oneil65,SS71}, according to which fast particles are able to pump the Langmuir spectrum, up to a saturation limit, resulting in trapping into the instantaneous potential well. The original treatment of the problem in \citer{OWM71} is particularly successful because it reduces the dynamics to a Hamiltonian $N$-body system, using dimensionless universal units. Such a model relies on the hypothesis that the beam propagates in a cold background plasma, feeling its response in the form of a real dielectric and treating the system as collisionless, \emph{de facto} neglecting any other reciprocal back-reaction among these two components. Furthermore, the particle beam is assumed to be tenuous, \ie the ratio between the beam and the plasma densities is a small control parameter of the interaction. However, as discussed in Sec.\ref{newtok}, when the beam-plasma system (BPS) (say also the bump-on-tail instability) is implemented to describe the outgoing radial transport of fast ions interacting with Alfv\'en waves (for a precise mapping between the velocity to the radial space, see \citers{nceps,nceps19}), some of the approximations underlying the original model could result in being inadequate.

In this paper, we start by providing, in Sec.\ref{secp}, a detailed derivation of the Poisson equation at the ground of the dielectric approximation for the background plasma, based on a careful investigation of the instantaneous linear response of the bulk when it is crossed by the fast particles. Although the main ideas of the present derivation were implicitly contained in \citer{OWM71}, nonetheless the present contribution explicitly clarifies which approximations are required to summarize the plasma response in terms of a dielectric function. Such an analysis is of interest in order to better clarify the approach developed in Sec.\ref{sec3}, where the thermal plasma back-reaction is described via fluid dynamics, and therefore the emergence of a return current can be described.

Then, after providing in Sec.\ref{sec1} the standard $N$-body description of the BPS, in Sec.\ref{sec2} we investigate the details of the linear dispersion relation in the case of an initial positive slope of the beam distribution in the velocity space. In particular, we address non-perturbative effects and the breakdown of the inverse Landau damping expression where the ratio between the growth rate and the real frequency shift with respect to the plasma frequency becomes finite.

In Sec.\ref{sec3}, we analyse the problem concerning the induction of a return current within the background plasma, by coupling the $N$-body dynamics with the electrostatic fluctuations in the plasma, described via a linear fluid representation. We arrive at a revised model, in which the Poisson equation has memory of the presence of the Langmuir oscillations. The numerical implementations of this scheme, both for a single mode and for the so-called quasi-linear model (when many modes are included in a broad spectrum), are developed. As a result, we demonstrate that the spectrum and the beam distribution function morphology are similar to those predicted by the original treatment. However, now it is possible to calculate a return current in the background plasma, whose presence depresses the saturated spectrum energy and the mode linear growth rates. This is of clear importance in view of the realistic prediction for relaxed fast-particle profiles.

Finally, in Sec.\ref{sec4} we investigate the effect of friction on the motion of fast particles, in order to clarify how their distribution function and the spectrum are affected in the quasi-linear scenario. The friction we are addressing consists of the interaction of the fast particles with the bulk of the plasma charges, as integrated at a distance greater than the Debye scale. In other words, we neglect the effects of collisions along the fast-particle motion, because of their extreme dilution, but we account for collective interaction, acting as a force contrasting the particle dynamics, in analogy to what has already been studied in the literature for heavy ions \citep{NR55}. We find the interesting feature of a resonance detuning, able to significantly deform the particle distribution function and the mode spectrum on a given time of the non-linear dynamics. This effect is due to the loss of energy that the particles undergo under the collective friction of the plasma charges. Actually, a rather efficient displacement of particles takes place in the phase space towards the low-velocity region, which depresses those resonant modes at smaller wavelength. Clearly, after a sufficiently long time, the resonant detuning would result in a depletion of the spectrum energy, since all particles would essentially be no longer resonant.

\section{Derivation of the Poisson equation for the BPS}\label{secp}
The BPS \citep{OM68,OWM71} deals with a fast electron beam injected into a one-dimensional plasma, which is treated as a cold linear dielectric medium supporting longitudinal electrostatic Langmuir waves. This section is devoted to setting up the proper form of the Poisson equation for the electric field. Such a construction follows and clarifies the general ideas contained in \citer{OWM71}. 
In particular, we explicitly outline how the shift in frequency appearing in the electric susceptivity coincides with the modulation of the electric field amplitude induced by the interaction with the beam particles.

Let us now consider the distribution function in the velocity space $F^p(v)$ of the equilibrium one-dimensional thermal plasma, the associated perturbed plasma electron distribution function $\delta f^p(t,x,v)$ and the fast electron tenuous beam distribution function $f(t,x,v)$ (here $x$ is the coordinate of the one-dimensional periodic slab of plasma). The interaction between the plasma and the beam electrons is mediated by the electric field $E(x,t)$, according to the Poisson equation
\begin{equation}
\p_x E = - 4\upi e \int_{-\infty}^{\infty}\!\!\!\mathrm{d}v (f+ \delta f^p)\;,
\label{pre1}
\end{equation}
where $e$ denotes the electron charge. The equation above determines the electric field via the charge density of the external beam and the charge displacement induced in the thermal plasma. It is worth noting that the charge distribution associated with $F^p$ is canceled by the ion neutralizing background, whose dynamics are neglected completely in this treatment, due to the slow response of ion motion. Expanding in Fourier integral both $E$ and all the space-dependent distribution functions, the Poisson equation is written, for each wavenumber $k$, as
\begin{equation}
\mathrm{i}kE_k(t)=-4\upi e \int_{-\infty}^{\infty}\!\!\!\mathrm{d}v(f_k(t,v)+\delta f_k^p(t,v))\;.
\label{pre2}
\end{equation}
Neglecting the coupling between different (non-zero) $k$, $\delta f_k^p$ obeys the following reduced Vlasov equation
\begin{equation}
\p_t \delta f_k^p(t,v) + \mathrm{i}kv\delta f_k^p(t,v) = \frac{e}{m}E_k(t)\p_v ( F^p(v) + \delta f_0^p(t,v))\;.
\label{pre3} 
\end{equation}
Here, we considered only the contribution coming from the zeroth wavenumber because such homogeneous ($x$-independent) contribution is the only one containing the equilibrium distribution function derivative in the particle velocity. All of the other components $\delta f_k^p$ would provide an intrinsically non-linear contribution to the Vlasov equation.
If we additionally consider $|\delta f_0^p|\ll F^p$, this equation describes the linear response of the plasma as a dielectric structure. The solution of \eref{pre3} reads
\begin{equation}
\delta f_k^p(t,v) = \frac{e}{m} \int_0^t \!\!\!\mathrm{d}t^{\prime} E_k(t^{\prime}) \mathrm{e}^{\mathrm{i}kv(t^{\prime} - t)}\p_v (F^p(v)+\delta f_0^p(t',v))\;.
\label{pre4}
\end{equation}
Without loss of generality, we can use the following functional form for the electric field:
\begin{equation}
E_k(t)=\bar{E}_k \exp \Big[ -\mathrm{i}\int_0^t \!\!\!\mathrm{d}t^{\prime}\omega_k(t^{\prime}) \Big]\;,
\label{pre5}
\end{equation}
where $\bar{E}_k=const.$ As a natural consequence, we can also write
\begin{equation}
E_k(t^{\prime})=E_k(t)\exp \Big[-\mathrm{i}\int_t^{t^{\prime}}\!\!\!\mathrm{d}y\,\omega_k(y)\Big]\;.
\label{pre6}
\end{equation}
Using the general expression of the electric field above, \eref{pre4} can be rewritten as
\begin{equation}
\delta f_k^p(v,t) = \frac{e}{m} E_k(t) \int_0^t \mathrm{d}t^{\prime} \exp \Big[ \mathrm{i}kv(t^{\prime}-t)-\mathrm{i}\int_t^{t^{\prime}}\!\!\!\mathrm{d}y\, \omega_k(y)\Big]\,
\p_v (F^p(v)+\delta f_0^p(t',v))\;.
\label{pre7} 
\end{equation}
In this expression, in order to focus on resonant-driven Langmuir modes, we can set $kv\simeq\omega_k\simeq\omp$ \citep{OM68}, where $\omp=\sqrt{4\upi n_p e^2/m}$ denotes the plasma frequency ($m$ is the electron mass). Hence we explicitly set $\omega_k=\omega_p+\delta \omega_k$, where $\delta\omega_k$ is a small complex function. We observe that the integral in $t^{\prime}$ contains only one rapidly oscillating term $\mathrm{e}^{-\mathrm{i}kv(t^{\prime}-t)}\simeq \mathrm{e}^{-\mathrm{i}\omega_p(t^{\prime}-t)}$ multiplied by slow varying terms: $F^p$ does not depend on time and, for a cold plasma, it is peaked around $v\simeq 0$, while $\delta f_0^p$ is expected to have a typical time scale much larger than $1/\omp$. Thus, we can bring out of the integral in $t^{\prime}$ the term in the distribution function derivative. Moreover, the time integral in $t^{\prime}$ is dominated by the values taken in its extremes (where the rapidly oscillating term does not phase mix), especially at $t=t^{\prime}$ where one has no damping due to the imaginary part of $\delta \omega_k$: there, the integral in $y$ gives almost $\omega_k(t)(t^{\prime}-t)$ (see the appendix in \citer{OWM71}). We finally get
\begin{equation} 
\delta f_k^p(v,t) = \frac{e}{m}E_k(t) \frac{\p_v (F^p(v)+\delta f_0^p(t,v))}{\mathrm{i}(kv-\omega_k(t))}\;.
\label{pre7bis}
\end{equation}
Substituting this expression in the Poisson equation \reff{pre2}, we obtain
\begin{align}
&\mathrm{i}kE_k(t)\Big[1-\frac{4\upi e^2}{mk}\int_{-\infty}^{\infty}\!\!\!\mathrm{d}v \frac{\p_v F^p(v)}{kv-\omega_k(t)} \Big]\equiv\nonumber\\[5pt]
&\qquad\equiv \mathrm{i}kE_k(t)\epsilon(\omp+\delta \omega _k(t), k) =\nonumber\\
&\qquad=-4\upi e\Big[ \int_{-\infty}^{\infty}\!\!\!\mathrm{d}v f_k(t,v)\Big] - \mathrm{i}kE_k(t) \delta \epsilon (t,k)\;,
\label{pre8}
\end{align}
where $\epsilon(\omp+\delta \omega _k(t), k)$ is the instantaneous (linear) dielectric function, while the correction $\delta \epsilon (t,k)$ is defined as
\begin{equation} 
\delta \epsilon \equiv - \frac{4\upi e^2}{mk} 
\int_{-\infty}^{\infty}\!\!\mathrm{d}v \frac{\p_v\delta f_0^p(t,v)}{kv - \omega_k(t)}\;.
\label{pre9} 
\end{equation}

In the case of a cold plasma, \ie $\epsilon=1-(\omega _p/\omega)^2$, for a Langmuir wave ($\omega_k\simeq\omp$), we immediately get the relation 
\begin{align}
\mathrm{i}k\epsilon E_k(t) \simeq \mathrm{i}k\frac{2}{\omega_p}\delta \omega _k(t) E_k(t)
= -\frac{2k}{\omega _p}\p_t E_k(t) - 2\mathrm{i}kE_k(t)\;,
\label{pre10}
\end{align} 
where we made use of the representation \reff{pre5} for the electric field. Finally, if we neglect the small non-linear contribution $\delta \varepsilon$, the Poisson equation takes the well-known form
\begin{align}
\p_t E_k(t) = - \mathrm{i} \omega _p E_k(t) + \frac{2\upi e\omega _p}{k}\int_{-\infty}^{+\infty}\!\!\!\mathrm{d}v f_k(t,v)\;.
\end{align}
In this derivation, we have outlined the relation existing between the function $\delta\omega_k$ appearing in \eref{pre5} and the effective frequency shift of the dielectric function.

\section{Hamiltonian description of the beam-plasma interaction}\label{sec1}
The BPS is here addressed by considering the basic assumption that the beam density $n_B$ is much smaller than that of the background plasma $n_p$. We adopt the Hamiltonian $N$-body formulation of the problem described in \citers{CFMZJPP,ncentropy} and refs. therein, where the broad supra-thermal particle beam self-consistently evolves in the presence of $M$ modes at the plasma frequency, \ie $\omega_j\simeq\omp$ for $j=1,\,...,\,M$. This ensures that the dielectric function of the cold background plasma, \ie $\epsilon=1-\omp^2/\omega_j^2$, is nearly vanishing \citep{OM68} and allows casting the Poisson equation for each plasma oscillation into the form of a simple evolution equation. In this scheme, particle trajectories are solved from Newton's law \citep{OWM71}, while one single mode in the fluctuation spectrum is initially set in order to be resonantly excited (linearly unstable) with a wavenumber $k=\omega/v_r$, where $v_r$ is a given initial resonance velocity.

As already discussed, the one-dimensional cold plasma equilibrium is taken as a periodic slab of length $L$, and the position along the $x$ direction for each particle is now labelled by $x_i$ while $N$ denotes the total particle number ($i=1,\,...\,N$). Beam particle positions are scaled as $\bar{x}_i=x_i(2\upi/L)$, and the Langmuir wave scalar potential $\varphi(x,t)$ is expressed in terms of the Fourier components $\varphi_j(k_j,t)$ (where $k_j$ is the wavenumber). Introducing the parameter $\eta=n_B/n_p$, we use the following dimensionless variables: $\tau=t\omp$, $u_i=\bar{x}_i'=v_i(2\upi/L)/\omp$, $\ell_j=k_j(2\upi/L)^{-1}$ (integers), $\phi_j=(2\upi/L)^2 e\varphi_j/m\omp^2$ and $\bar{\phi}_j=\phi_j \mathrm{e}^{-\mathrm{i}\tau}$. The prime denotes $\tau$ derivative and barred frequencies and growth rates associated with the beam-plasma instability are normalized as $\bar{\omega}=\omega/\omp$ and $\bar{\gamma}=\gamma/\omp$, respectively.

The BPS is governed by the following system
\begin{subequations}\label{mainsys1}
\begin{align}
\bar{x}_i' &=u_i \;,\\
u_i' &=\sum_{j=1}^{M}\big(\mathrm{i}\,\ell_j\;\bar{\phi}_j\;\mathrm{e}^{\mathrm{i}\ell_j\bar{x}_{i}}+\mathrm{c.c.}\big)\;,\\[-8pt]
\bar{\phi}_j' &=-\mathrm{i}\bar{\phi}_j+\frac{\mathrm{i}\eta}{2\ell_j^2 N}\sum_{i=1}^{N} \mathrm{e}^{-\mathrm{i}\ell_j\bar{x}_{i}}\;.
\end{align}
\end{subequations}
For a single mode, we remark that resonance conditions are rewritten $\ell u_{r}=\bar{\omega}$. Moreover we assume that the warm beam is initialized in the velocity space with an assigned distribution function $F(u)$ (the time-dependent beam profile is denoted by $f_B(\tau,u)=f_0(\tau,u)/n_B$). In the simulations, we analyse a total of $N=10^{6}$ particles and we solve \erefs{mainsys1} using a Runge-Kutta (fourth-order) algorithm. The initialization in the velocity space is formal and, thus, particles are free to spread in this coordinate, while we set uniform initial conditions for particle positions in $[0,\,2\pi$] (we also apply periodic boundary conditions for $\bar{x}_i$ in this range). More specifically, we discretize the adopted initial beam profile in 500 delta-like beams each of them having a single fixed velocity. Each cold beam is initialized also with random generated positions and with a number of particles provided by the discretization of the considered initial profile (normalized to $N$). This provides the initial $2\times N$ array for $(\bar{x}_i,u_i)$, which is dynamically evolved in time self-consistently with the modes. Finally, the modes are initialized at $\mathcal{O}(10^{-12})$, in order to guarantee the initial linear stage of evolution. For the considered time scales and for an integration step $\Delta\tau=0.1$, as we discuss in detail in \S \ref{sec3}, both the total energy and momentum are safely conserved.

\subsection{Relevance of the BPS to the tokamak configuration}\label{newtok}
It is well known that the energetic particle (EP) excitations of global instabilities in burning plasmas and the corresponding non-linear transport are multi-scale processes \citep{ZCrmp} and, thus, reduced models for the profile relaxation are often considered in view of computational savings. When addressing multiple Alfv\'en eigenmodes (AEs), the quasi-linear model \citep{BB95b,BB96b,GG12} corresponds to the typical reduced description (a detailed comparison between reduced and fully non-linear schemes for turbulence-driven EP transport can be found in \citer{bourdelle16}). As already stated, in the group of reduced models, the BPS can be considered to face the analysis of the non-linear interactions between EPs and Alfv\'enic fluctuations \citep{BB90a,BB90b,BB90c,CZ13,ZC15njp}. In particular, a comprehensive study of the applications and specific restrictions is given in \citers{BS11,ZCrmp}.

In the following, with the aim of illustrating the direct implementation of the BPS as a paradigm for describing the burning plasma scenario, we introduce a mapping technique between the reduced radial profile of real systems and the velocity coordinate of the BPS. A validation of this approach can be found in \citer{nceps19} (see also \citer{nceps}), where the ITER $15\,$MA baseline scenario is addressed in the presence of the least damped 27 toroidal AEs, and also in \citer{bnc} for an application to the case of EP-driven geodesic acoustic modes.

The relation between the independent variables is a linear correspondence which preserves the non-linear EP profile redistribution. In particular, the AE/EP scheme must be dimensionally reduced (using standard averaged distribution functions) in order to describe the dynamics as a one-dimensional non-autonomous system \citep{BW14pop}. The map can be defined for a fixed single resonance, starting from the corresponding resonance condition:
\begin{align}
\bar{\omega}^{\aes}(s)-\bar{\omega}^{\aes}(s_r)=\Omega\, \ell(u-u_r)\;,
\end{align}
where $\Omega$ is a normalization constant to be determined during the procedure, and we use standard nomenclature for the EP/AE scenario\footnote{Here, $s=r/a$ is the normalized tokamak radius (with $a$ the minor radius) and the AE frequency reads $\omega^{\aes}$ and it is normalized (barred quantity, valid also for the related growth rate) in $\omega_{A0}$ units ($\omega_{A0}=v_{A0}/R_0$, where the Alfv\'en speed at the magnetic axis is dubbed $v_{A0}$ and $R_0$ denotes major radius). As before, the subscript $r$ represents the resonance value of a quantity.}. This mapping procedure is defined as local, by using the expansion $\bar{\omega}^{\aes}(s)=\bar{\omega}^{\aes}(s_r)+(s-s_r)\p_s\bar{\omega}^{\aes}$, which is always valid when the frequency results in a sufficiently smooth function of the radius over the resonance region. The procedure can be finalized by implementing the suitable boundary conditions $s=0\;\mapsto\;u=u_{max}$ and $s=1\;\mapsto\;u=0$, and by introducing a parameter $\ell_1$ which substantially fixes the arbitrary periodicity length $L$ for the BPS one-dimensional slab (cf. the normalization introduced above) and it corresponds to the minimum mode number that can be analysed. The closed linear link between the radial profile and the velocity coordinate reads now
\begin{align}
u=(1-s)/\ell_1\;.
\end{align}
In this scheme, multiple modes can be introduced in the dynamics simply by considering the corresponding resonance conditions written in the radial space: $\ell_{j}=\ell_1/(1-s_{rj})$. A real predictive comparison of the two systems can be implemented once the density parameter of the BPS, \ie $\eta$, is determined. Without entering the details of this step, $\eta$ can be evaluated by assuming a direct proportionality between the linear growth rates (normalized to the mode frequency) of the BPS and the AE/EP scenario. The proportionality constant can be fixed requiring that the non-linear transport is preserved \citep{bnc,nceps19}. Then, the integration of the linear dispersion relation for the BPS, which is analysed in the next section, directly provides the value of the density parameter.

This analysis underlines how the BPS can be addressed as a paradigm for describing specific burning plasma scenarios. The powerful features of this approach can be seen in the linear link between $s$ and $u$, and in the very reduced computational demand of the BPS. In this respect, in the following sections, we analyse some specific aspects of the beam-plasma instability having in mind their direct outcomes when implemented in the radial profile of real systems.

\section{Linear dispersion relation: non-perturbative effects}\label{sec2}

In this section, we review go into more depth of some specific features of the dispersion relation. The following analysis is valid on a purely linear regime, having the aim of clarifying the non-local character of the distribution function in the velocity space.

The dispersion relation, in correspondence to electric field perturbations $\propto \mathrm{e}^{\mathrm{i}(kx-\omega t)}$, is formally written \citep{OM68,LP81}
\begin{align}
1-\frac{\omp^2}{\omega^2}=\frac{\eta \omp^2}{k^2}
\int_{-\infty}^{+\infty}\!\!\!\!\!\mathrm{d}v\,\frac{k\p_v \hat{F}(v)}{k v-\omega}\;,
\end{align}
where the left hand side corresponds to the cold plasma dielectric function $\epsilon(\omega)$ and $\hat{F}(v)$ is the normalized (to unity) initial beam distribution function in the velocity space.

Since the dielectric is expanded near $\omega\simeq\omp$ in \erefs{mainsys1} \citep{OWM71,CFMZJPP}, the left hand side of the dispersion relation must be consistently rewritten as $\epsilon\simeq2(\bar{\omega}-1)$. In order to study the instability features of one single resonant mode having mode number $\ell$, we explicitly write $\bar{\omega}=\bar{\omega}_0+\mathrm{i}\bar{\gamma}_L$, where $\bar{\omega}_0$ contains a small real frequency shift with respect to the Langmuir mode frequency $\omp$ (the resonance velocity is, thus, $u_r=\bar{\omega}_0/\ell\simeq 1/\ell$) and $\bar{\gamma}_L$ denotes the normalized linear growth rate. Using the proper normalization, we can finally rewrite the dispersion relation as
\begin{align}\label{disrel}
2(\bar{\omega}_0+\mathrm{i}\bar{\gamma}_L-1)-\frac{\eta}{\ell M}
\int_{-\infty}^{+\infty}\!\!\!\!\!\!\!\mathrm{d}u\,\frac{\p_u F(u)}{\ell u-\bar{\omega}_0-\mathrm{i}\bar{\gamma}_L}=0\;,
\end{align}
where we transformed $\hat{F}(v)\to F(u)$ with $M=\int \mathrm{d}u F(u)$.

\eref{disrel} is numerically integrated for different mode number by fixing the parameter $\eta$ and, as initial distribution function, we consider a half-Gaussian positive slope $F(u)=0.5\;\textrm{Erfc}[a-b\,u]$ (with $a\simeq6.8$ and $b\simeq4537$).
\begin{figure}
\centering
\includegraphics[width=0.45\textwidth]{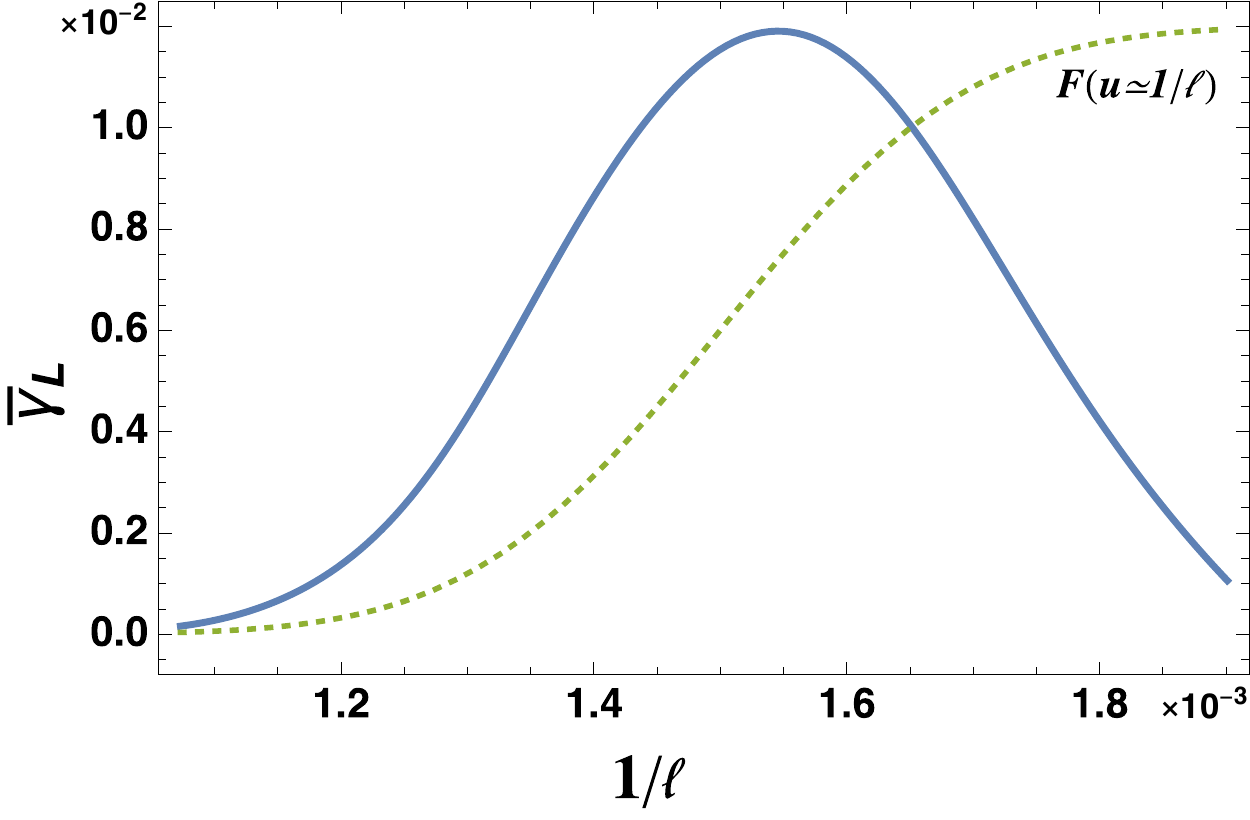}\hfill
\includegraphics[width=0.46\textwidth]{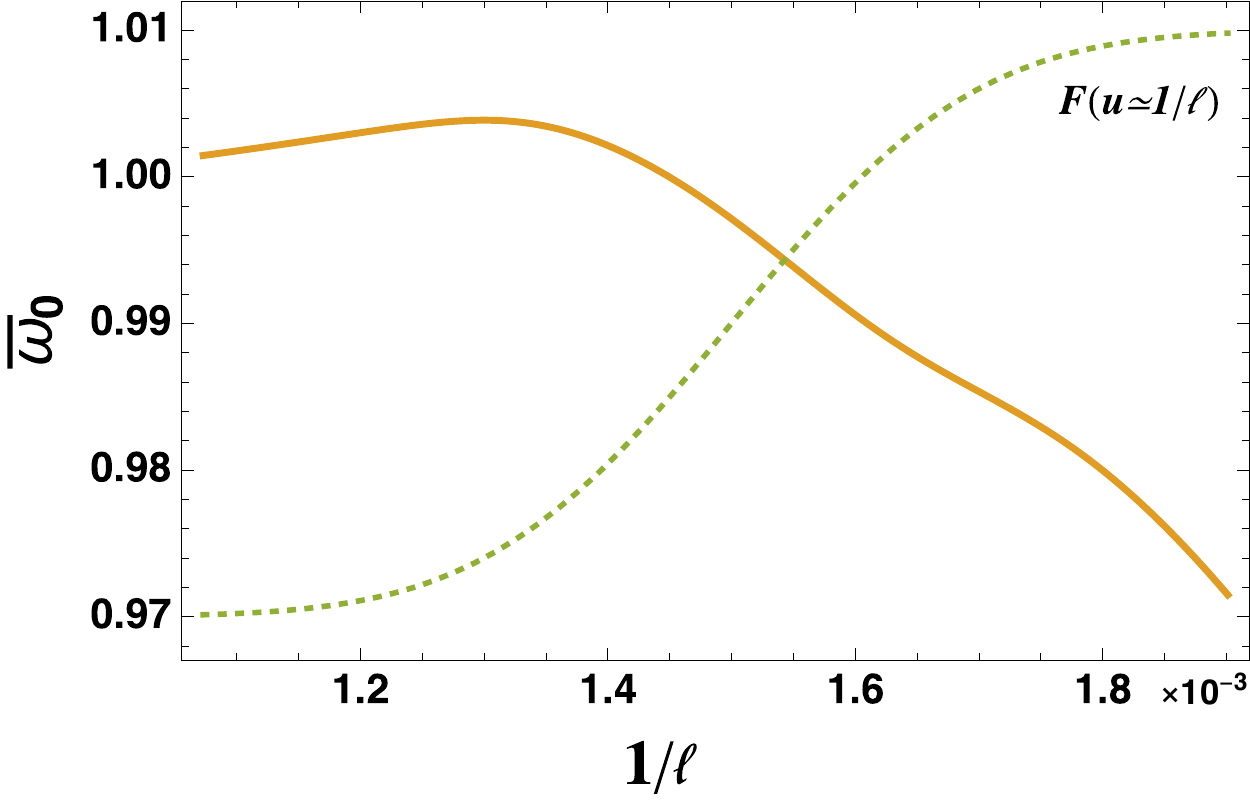}
\caption{Plot of the linear growth rate $\bar{\gamma}_L$ (left panel) and of the real part of the frequency $\bar{\omega}_0$ (right panel), as a function of the mode-number, obtained numerically by integrating \eref{disrel} for the reference case $\eta=0.0007$. Green dashed line qualitatively depicts (off-scale) the initial distribution function represented assuming $F(u=1/\ell)$.
\label{fig_glom}}
\end{figure}
The corresponding resonant velocities occur in different regions of the distribution function varying from $u_r\simeq0.0011$ to $u_r\simeq0.0019$. From the numerical integration plotted in \figref{fig_glom} and specified for a reference case $\eta=0.0007$, it emerges how growth rate values increase as the associated resonant velocity approaches the inflection point of $F(u)$, where, of course, the drive of the inverse Landau damping $\p_u F$ is maximum. Furthermore, the real part of the frequency $\bar{\omega}_0$ starts to deviate from unity on the right-hand flat profile of the distribution function. Actually, in that region the local value of $\p_u F$ is essentially negligible and a finite growth rate (necessarily associated with a real frequency shift with respect to the plasma frequency) marks the breakdown of the perturbative Landau damping expression.

Let us now compare the obtained results with respect to a standard analytical and a semi-analytical integration of the dispersion relation. By linearizing \eref{disrel} fixing $\bar{\omega}_0=1$, one can get the well-know Landau formula for the linear growth rate (here denoted by the subscript lin) \citep{LP81}:
\begin{align}\label{drlin}
\bar{\gamma}_{lin}=\frac{\upi\eta}{2\ell^2 M}\p_u F\big|_{u=1/\ell}\;.
\end{align}
Meanwhile, the dispersion relation can be also analytically integrated in terms of the Faddeeva function $w(x)$ and residue contributions \citep{conte61}. We easily get $\p_u F=A\,\mathrm{exp}[-B^2(C-Du)^{2}]$, and \eref{disrel} provides the following expression (specified using the subscript $F$) \citep{malik16}:
\begin{align}\label{drfad}
2(\bar{\omega}_{0F}+\mathrm{i}\bar{\gamma}_F-1)-\frac{\eta A}{\ell M}\;\mathrm{i}\upi(2\mathrm{e}^{-X}-w(X))=0\;,
\end{align}
with $X=BC-BD(\bar{\omega}_{0F}+\mathrm{i}\bar{\gamma}_F)/\ell$ (for the sake of completeness, in our specific case, we have: $A=2559.88$, $B=5.89$, $C=1.15$ and $D=770.0$), whose roots can be numerically determined.
\begin{figure}
\centering
\includegraphics[width=\dime]{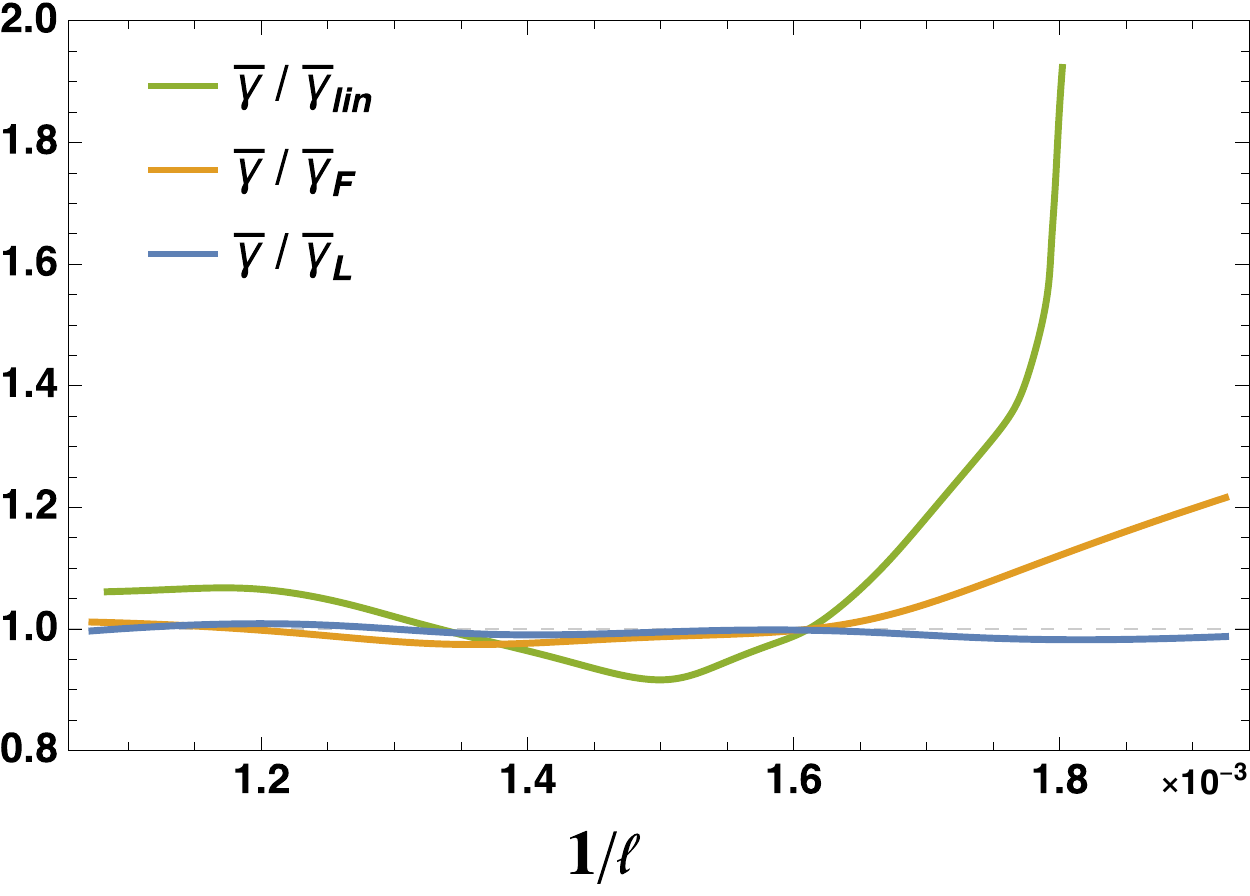}
\caption{Ratios between the growth rate $\bar{\gamma}$ (fitted from simulations) and $\bar{\gamma}_L$, $\bar{\gamma}_{lin}$ and $\bar{\gamma}_F$ evaluated from \erefs{disrel}, \reff{drlin} and \reff{drfad}, respectively, for different mode numbers ($\eta=0.0007$).
\label{fig_gamma}}
\end{figure}

In \figref{fig_gamma}, we plot the ratios between the growth rates $\bar{\gamma}$ fitted from the time behavior of single-mode simulations of \erefs{mainsys1} and the obtained values of $\bar{\gamma}_L$, $\bar{\gamma}_{lin}$ and $\bar{\gamma}_F$ from \erefs{disrel}, \reff{drlin} and \reff{drfad}, respectively, as a function of the analysed mode number for $\eta=0.0007$. While the perturbative inverse Landau damping expression can deviate up to $100\%$, the predicted $\bar{\gamma}_F$ values from \eref{drfad} appear more precise; nonetheless an underestimation of the simulated growth rate up to $20\%$ emerges in correspondence with the associated $u_r$ larger than the inflection point. The $\bar{\gamma}_L$ values calculated from \eref{disrel} almost perfectly match the numerical simulations. 

It is worth nothing that the considered linear problem actually contains a subtle issue concerning the non-perturbative character of the dispersion relation. The integral over the whole distribution function and the non-linear dependence upon the frequency make the estimate of the growth rate and the real frequency shift with respect to the plasma frequency intrinsically non-local. In particular, focusing on a region close to the inflection point of the distribution, the two methods giving $\bar{\gamma}_L$ and $\bar{\gamma}_F$, respectively, almost coincide, while the perturbative inverse Landau damping growth rate $\bar{\gamma}_{lin}$ overestimates the simulation results by about $10\%$. This is due to the fact that the residue and the principal value contribution are of the same order in this region.

We conclude by observing that, when addressing the Landau growth rate expression \eref{drlin}, we evaluate the quantity $\p_u F$ assuming the resonance at $u=1/\ell$, \ie we impose $\bar{\omega}_0=1$ ($\omega = \omega_p$) without considering the real frequency shift due to the presence of the beam. Actually, when the beam is tenuous (as in the considered case), we are near the marginal stability corresponding to $\bar{\gamma}_L\ll 1$ and $\bar{\omega}_0\simeq1$. Thus, neglecting the frequency shift when evaluating \eref{drlin} is expected to be a good approximation up to the considered order in $\bar{\gamma}$. This is certainly true when we consider modes resonant with velocities for which $\p_u F$ is near a maximum value, \ie near the inflection point of the distribution function. In fact, if we calculate the correction due to the real frequency shift, it is of order $\p_u^2 F|_{u=1/\ell} (\bar{\omega}_0-1)$, being clearly of higher order in the addressed velocity region. This allows us to consider the assumption $\bar{\omega}_0=1$ as a closed and simple prescription to describe the inverse Landau damping associated with the presence of a tenuous beam.

However, since we are investigating the dispersion relation in the flat regions of the distribution function (where $\p_u F$ is very small but $\p_u^2 F$ is finite), it becomes important to evaluate the modification of the Landau growth rate when the real frequency shift is taken into account as directly evaluated from the exact dispersion relation \eref{disrel}.
\begin{figure}
\centering
\includegraphics[width=\dime]{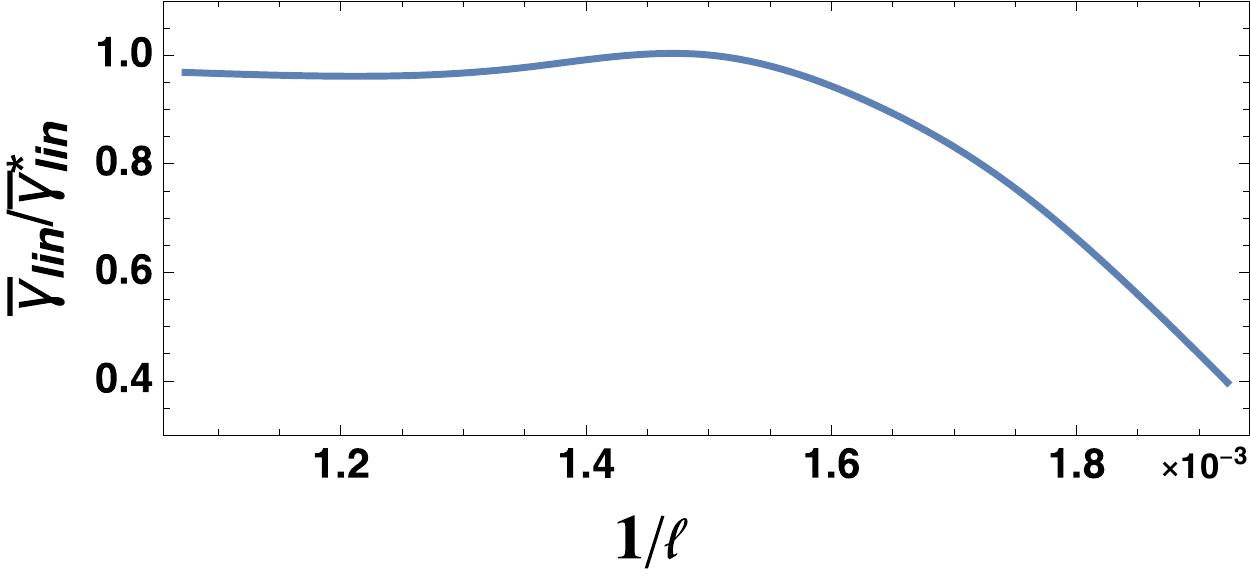}
\caption{Ratio between $\bar{\gamma}_{lin}$ using $\bar{\omega}_0=1$ and $\bar{\gamma}_{lin}^*$ evaluated considering $\p_u F|_{u=\bar{\omega}_0/\ell}$ in \eref{drlin}, where $\bar{\omega}_0$ are plotted in \figref{fig_glom} ($\eta=0.0007$).
\label{fig_gamma_star}}
\end{figure}
In \figref{fig_gamma_star}, we plot the ratio between $\bar{\gamma}_{lin}$ calculated using $\bar{\omega}_0=1$ and that (dubbed $\bar{\gamma}_{lin}^*$) determined using the numerical frequency shift, \ie by considering $\p_u F|_{u=\bar{\omega}_0/\ell}$ in \eref{drlin}. It is immediately evident that this ratio is $\lesssim1$ and the real frequency shift contribution is, thus, unable to eliminate the discrepancy with respect to the fitted growth rates from the simulation. This fact is significant because it guarantees that such observed discrepancy is really due to the non-local contributions coming to the whole profile of the distribution function, which can never be mimicked by a locally calculated growth rate. It is just the relevant effect of the non-local nature of the dispersion relation, the main issue of the present analysis, deepening the pioneering studies in \citers{OM68}; see also \citer{FG61} for a study of the dispersion relation in a hot plasma (in this latter work, the existence was outlined of a strongly damped electronic acoustic mode).

\section{Return current in the beam-plasma instability}\label{sec3}

As already discussed, in the original treatment the plasma is considered as a linear dielectric. The validity of this assumption relies on the condition $\eta^{1/3}\ll1$, inasmuch as non-linear terms of the expansion of the background charge density are shown to give higher-order contributions to the Poisson equation (see Appendix in \citer{OM68}).

In this section, we provide a derivation of the BPS dynamics based on the description of the background plasma as a fully ionized fluid. In the magnetohydrodynamic scheme, its density and velocity are governed by the continuity and Euler equations, respectively. In the latter, we neglect any dissipative effects, such as friction, which we will separately examine in the next section. According to the assumption of a cold plasma, we also neglect the pressure gradient by considering only the electrostatic interaction expressed by the Poisson equation.

Although we consider only linear perturbations to the background plasma, as in the linear dielectric approach, the fluid representation allows us to give a quantitative estimate of the emerging non-zero current in the plasma, consisting of a combination of two effects: a fast oscillation of the electrons at the plasma frequency $\omp$, having a small amplitude and essentially non-relevant impact on the long-time-scale dynamics; and a return current in the opposite direction to that associated with the beam dynamics. The return current evolution closely resembles that one of the electric field: after an initial exponential growing phase, its amplitude saturates and starts oscillating around a non-zero value at the bounce frequency of the plasma electrons in the potential well. Its inclusion affects the growth rate of the electric field instability, even more markedly in the multi-mode case, where also the saturation levels are systematically lower.

In the following, $\bar{n}_e$ and $\bar{n}_B$ denote the plasma electron and the beam densities, respectively, and $u_e$ characterizes the plasma electron fluid velocity. Barred densities are normalized in units of the uniform density $n_p$ of the neutralizing ion background, assumed still.

In this setting, the equations describing the BPS take the form:
\begin{subequations}\label{eq-matt-1}
\begin{align}
\bar{n}_e' + \partial_{\bar{x}}\left(\bar{n}_e\,u_e\right) = 0\;,\label{eq-matt-1-1}\\
u_e' + u_e\partial_{\bar{x}}u_e - \partial_{\bar{x}}\phi=0 \;,\\
-\partial^2_{\bar{x}}\phi-1+\bar{n}_e+\bar{n}_B=0\;.\label{hgfiughdfsiug}
\end{align}
\end{subequations}
We now look for small perturbations in the plasma electron density and velocity, indicated by subscript ``1'', \emph{i.e.} $\bar{n}_e=1+\bar{n}_1$ and $u_e=0+u_1$. Substituting these expressions in the system above, and retaining only first-order terms, we find the following relation between $\bar{n}_1$ and $\bar{n}_B$:
\begin{align}\label{eq-matt-3}
(\partial^2_{\tau} + 1)\,\bar{n}_1=-\bar{n}_B\,,
\end{align}
whose formal solution is given by
\begin{align}\label{eq-matt-41}
\bar{n}_1=\tilde{n}_{1} - \int_0^\tau\!\!\! \mathrm{d}\tau' \left(\sin\left(\tau-\tau'\right) \,\bar{n}_B(\bar{x},\tau')\right)\,,
\end{align}
where $\tilde{n}_{1}=A\cos(\tau+\zeta)$ is the solution of the associated homogeneous equation, corresponding to fast Langmuir oscillations at frequency $\omp$. In the numerical simulations, we will assume the amplitude $A$ to be spatially constant and random phase $\zeta \in [0,2\upi)$. The integral term in \eref{eq-matt-41} can be rewritten as
\begin{align}\label{eq-matt-4}
-\bar{n}_B(\bar{x},\tau)+\eta\cos(\tau)+\int_0^\tau \!\!\! \mathrm{d}\tau' \left( \cos\left(\tau-\tau'\right) \,\partial_{\tau^\prime}\bar{n}_B(\bar{x},\tau')\right)\,.
\end{align}
Let us analyse the physical meaning of the expression above. The first term is an instantaneous neutralization of the net negative charge introduced by the beam in the system. The second term corresponds to a fast oscillation at the plasma frequency, with amplitude directly proportional to the density ratio $\eta$, which is a small parameter in our analysis. Finally, the integral term contains the non-trivial interaction between the two densities (its behavior will be characterized using numerical simulations at the end of the section).

Considering now \eref{eq-matt-1-1}, still retaining only linear terms, we find the following expression for the velocity perturbation:
\begin{align}\label{eq-matt-51}
u_1= \int\! \mathrm{d}\bar{x} \Big[ A\sin(\tau+\zeta)+\int_0^\tau\!\!\! \mathrm{d}\tau^\prime\left(\cos(\tau-\tau^\prime)\,\bar{n}_B(\bar{x},\tau^\prime)\right)\Big]\,,
\end{align}
which also provides, with opposite sign, the first-order plasma current defined as $\bar{J}_1=J_1/e=-\bar{n}_e u_e = -u_1$. As stressed above, also this velocity is a combination of a rapidly oscillating term, coming from Langmuir waves and giving no net effect, and an integral term akin to that encountered in \eref{eq-matt-4}. The latter gives a net contribution that can be identified as a return current which is directed in the opposite direction to the stream of the beam electrons.

Let us now perform a numerical study implementing the same approach of \citer{OWM71}, as in \S \ref{sec1}. More specifically, we consider the beam density as a sum of particles at coordinates $\bar{x}_i$. By Fourier transforming the space variable, we obtain
\begin{align}\label{eq-matt-nb}
\bar{n}_B(\bar{x},\tau)=\frac{\eta}{N}\sum_{j=0}^{M}\sum_{i=1}^N \exp(\mathrm{i}\ell_j\bar{x}_i)\;,
\end{align}
where we recall that $N$ is the number of particles and $M$ the number of Fourier modes. The final system reads
\begin{subequations}\label{eq-matt-5}
\begin{align}
\bar{x}_i'&=u_i \;,\\
u_i'&=\sum_{j=1}^{M}\big(\mathrm{i}\,\ell_j\;\phi_j\;\mathrm{e}^{\mathrm{i}\ell_j\bar{x}_{i}}+c.c.\big) \;,\label{eq-matt-5-1}\\
\phi_j &=\frac{1}{\ell_j^2}\Big( \tilde{n}_{1j} + \frac{\eta}{N}\sum_{i=1}^N \mathrm{e}^{-\mathrm{i}\ell_j\bar{x}_i} + \frac{\mathrm{i}\eta}{2N}\left(g^{+}_{j}-g^{-}_{j}\right) \Big) \;,\label{eq-matt-5-2}\\
\bar{J}_{1j} &= -\frac{\mathrm{i}}{\ell_j} \big[ A\sin(\tau+\zeta)-\frac{\eta}{2N}\left(g^+_{j}+g^-_{j}\right)\big]\;,\label{matt6}\\
\left(g^{\pm}_{j}\right)' &= \pm \mathrm{i} g^{\pm}_{j} + \frac{\eta}{N}\sum_{i=1}^N \mathrm{e}^{-\mathrm{i}\ell_j\bar{x}_i} \;,
\end{align}
\end{subequations}
where we underline that the final expression of the Poisson equation, \eref{eq-matt-5-2}, is algebraic, and we have introduced the two auxiliary variables $g^\pm$ simply to obtain a fully differential system.

\begin{figure}
\centering
\includegraphics[width=0.45\textwidth]{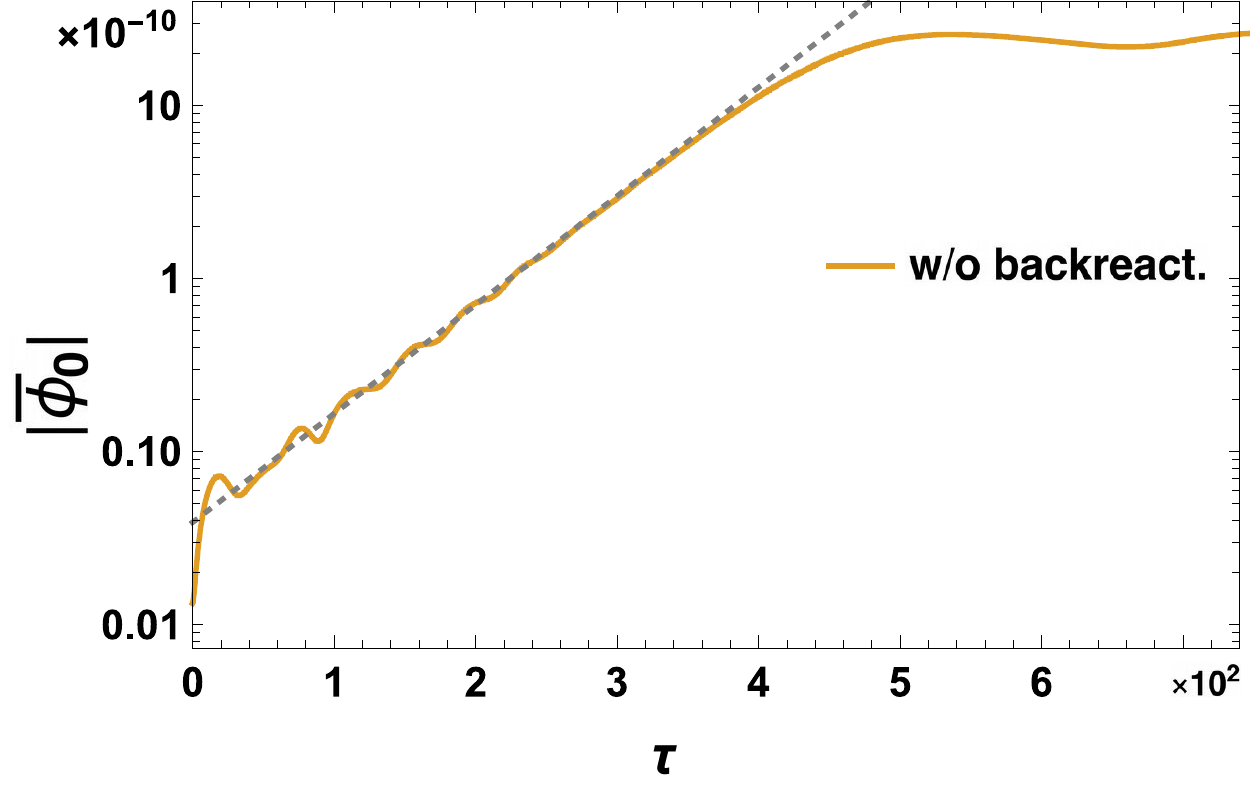}\hfill
\includegraphics[width=0.45\textwidth]{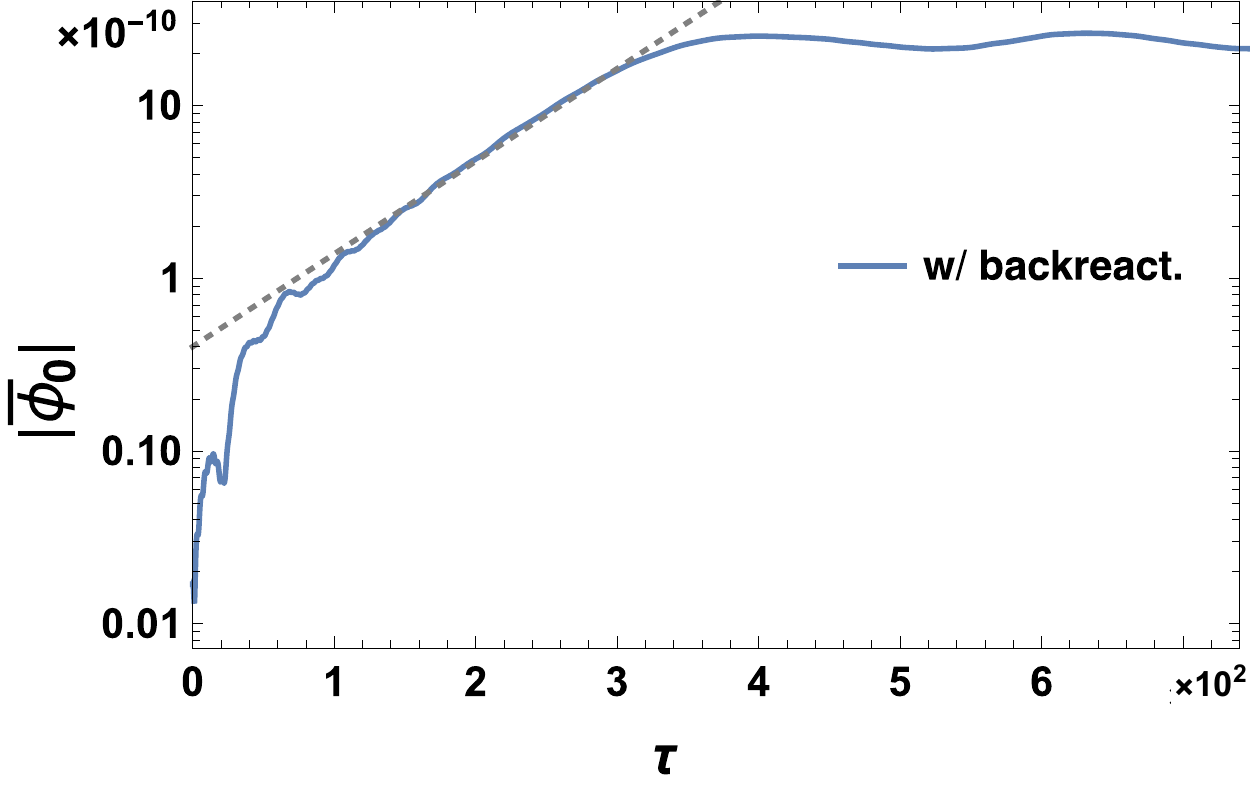}
\caption{Single-mode simulation: temporal evolution of the electric potential modulus in the case without (left panel) and with (right panel) back-reaction in logarithmic scale. Dashed lines correspond to the numerical fit for the linear phase.}
\label{fig-matt-1}
\end{figure}
\begin{figure}
\centering
\begin{minipage}{0.47\textwidth}
\centering
\vspace{-2pt}
\includegraphics[width=0.99\textwidth]{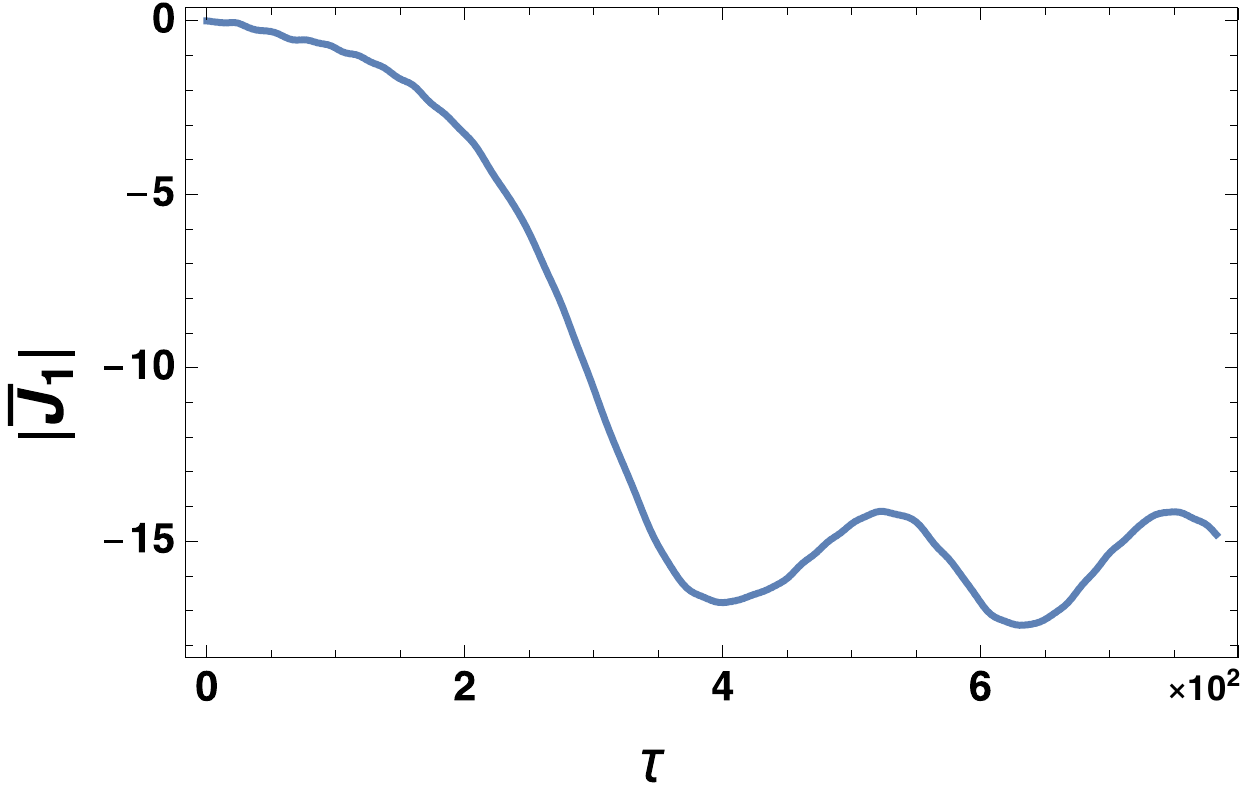}
\caption{Single-mode simulation: temporal evolution of the first-order current perturbation (equation \eref{matt6}) induced in the plasma ($\eta=0.0007$).}
\label{fig-matt-2}
\end{minipage}\hfill
\begin{minipage}{0.46\textwidth}
\centering
\includegraphics[width=0.99\textwidth]{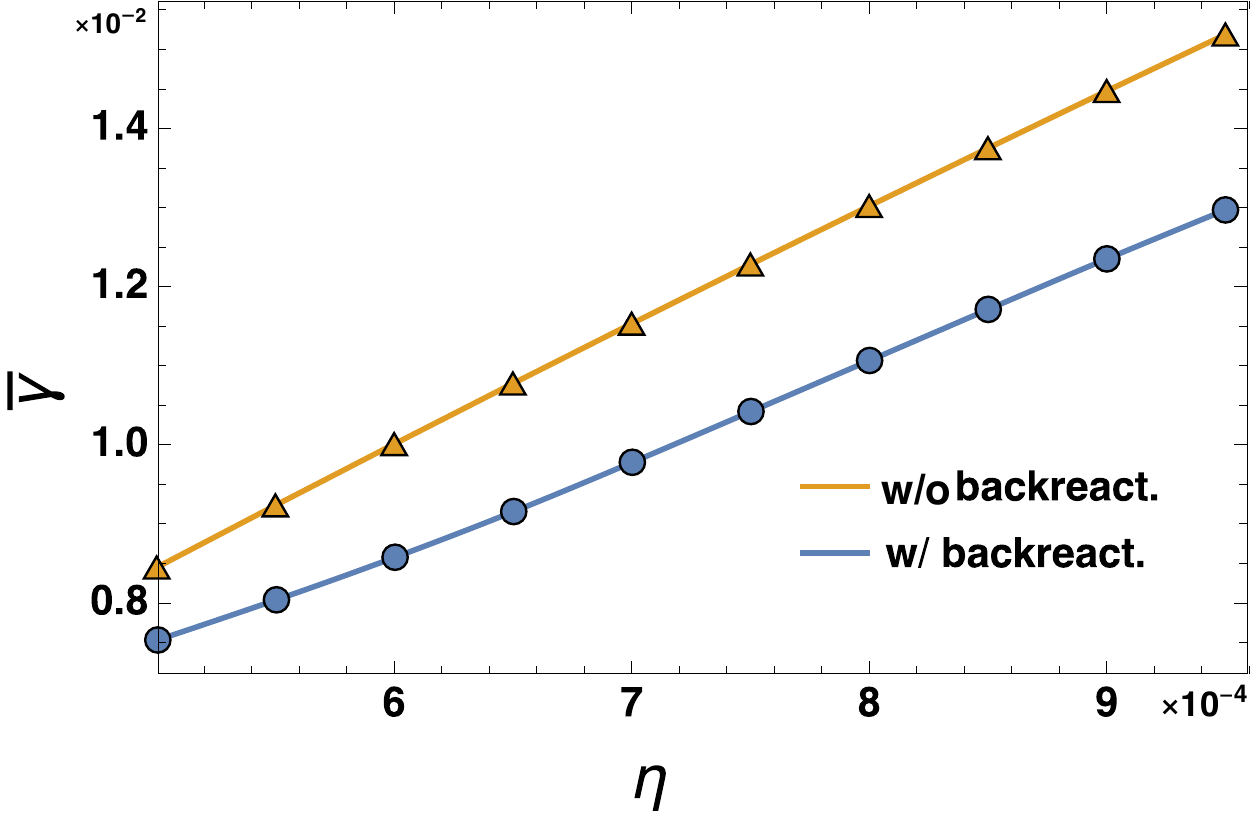}
\caption{Single-mode simulation: growth rate with and without back-reaction (specified in the plot) as a function of the density parameter $\eta$.}
\label{fig-matt-4b}
\end{minipage}
\end{figure}

As a first step, we study the single-mode case setting $\ell_1=666$ (associated with a resonant velocity $u_r\simeq1.5\times10^{-3}$) as the only mode included in the dynamics and by initializing the particles with the positive slope $F(u)$ introduced in the previous section (also the density parameter $\eta$ is fixed as 0.0007). The temporal evolution with and without back-reaction (\ie \erefs{mainsys1}) of the electric potential is shown in \figref{fig-matt-1}, while the current perturbation is plotted in \figref{fig-matt-2}.

\begin{figure}
\centering
\includegraphics[width=\dime]{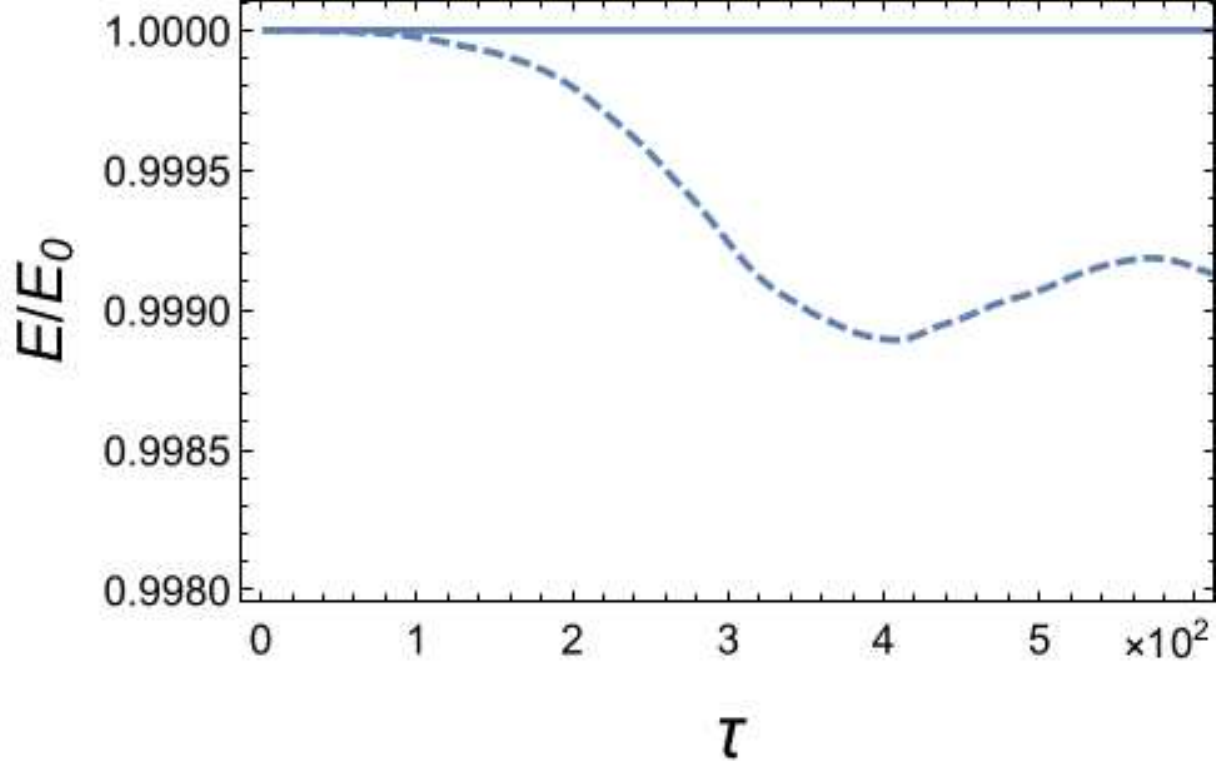}
\caption{Single-mode simulation: evolution of the total energy $E$ from \eref{jjjnnh}, normalized to its initial value $E_0$, with (solid line) and without (dashed line) the contribution of the background plasma current.}
\label{fig-matt-en}
\end{figure}

As it emerges from \figref{fig-matt-1}, after a longer pre-linear stage, the standard linear regime is clearly recovered and followed by mode saturation (non-linear phase) and oscillation. The value of the field at saturation results in being slightly lower if the back-reaction is on (for the analysed setup, it is lower by a factor of about $0.97$), indicating how part of the total energy is spent on the excitation of the plasma velocity perturbation. More specifically, the total energy of the system in the presence of back-reaction can be written as follows:
\begin{equation}\label{jjjnnh}
E=\frac{1}{2N}\sum_{i=1}^Nu_i^2+\sum_{j=1}^M\frac{\ell_j^2}{\eta}\left|\phi_j\right|^2+\sum_{j=1}^M\frac{1}{\eta}\left|\bar{J}_{1j}\right|^2 = const\;, 
\end{equation}
where we stress that the first two terms of this expression correspond to the total energy of the system without back-reaction \citep{OWM71}. In that case, the energy could only be exchanged between beam electrons and the electric potential, while now the third term proportional to $\bar{J}_1$ enters the dynamics. This corresponds to the energy of the excited plasma back-reaction current, whose incidence on the total energy conservation is shown in \figref{fig-matt-en}.

To further analyse the energy exchange between the plasma and beam electrons, in \figref{fig-matt-4b} we plot the fitted linear growth rates in the single-mode dynamics as a function of the density parameter $\eta$. Of course, in the case with back-reaction, the obtained growth rates are always below the standard ones leading to relative errors (for the analysed cases) of the order of $15-20\%$.
Since the linear growth rate is proportional to the non-linear velocity spread of beam particles \citep{nceps18}, the outlined behavior underlines the importance of the return current for the transport properties of the system and the resulting relaxed beam profiles.

\begin{figure}
\centering
\includegraphics[width=\dime]{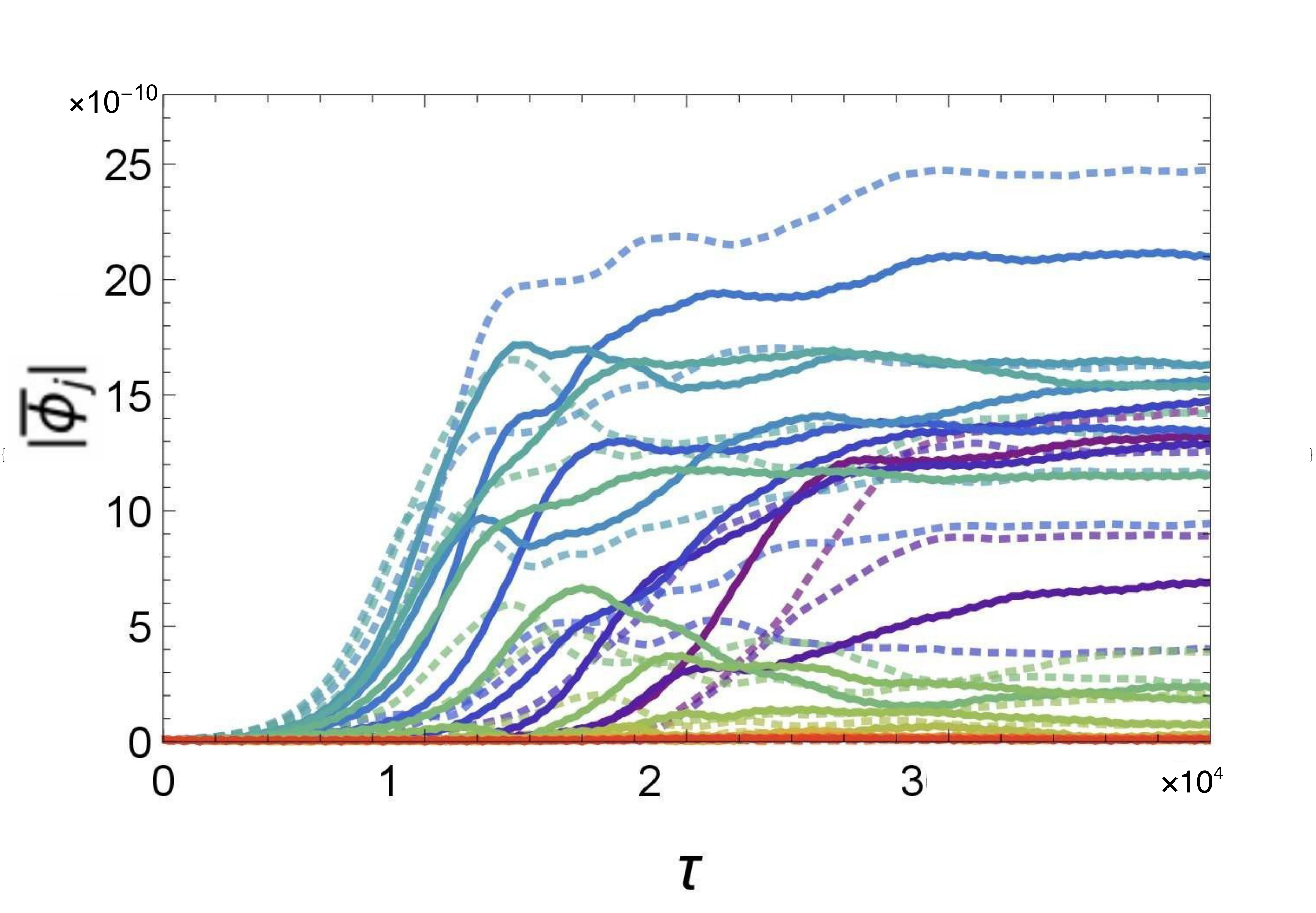}
\caption{Multi-mode simulation: mode evolution in the case with (solid lines) and without (dashed lines) back-reaction. We follow a standard rainbow colour scheme from $\ell_{min}=560$ (red) to $\ell_{max}=640$ (blue).}
\label{fig-matt-3}
\end{figure}

Let us now study the case of a broad beam initialized with a full Gaussian $F(u)=\exp(-(u-\bar{u})^2/2\sigma)/\sqrt{2\upi\sigma}$, with $\bar{u}=1/600$ and $\sigma=5.67\times10^{-5}$. We address $M=21$ modes chosen to resonate in the interval $0.00156\leqslant u \leqslant 0.00178$, \ie $\ell_{min}=560$ and $\ell_{max}=640$, having equispaced resonant velocities ($\eta=0.0007$). With the chosen set-up, the particle velocity spreads associated with the mode saturation allow the beam particles to move from one resonance to the other. Each mode, due to the limited width of the velocity spread, overlaps with its neighbors allowing a cascade process of diffusive kind, \ie the so-called quasi-linear dynamics (cf. \citer{ncentropy} and references therein for a Hamiltonian treatment of this process, in which the back-reaction is neglected). The evolution of the electric potentials is shown in \figref{fig-matt-3}, where we also report the case without back-reaction as reference.
The obtained results confirm the previous analysis outlining the larger energy exchange between particles and modes than in the standard case (here, it is evident also in the saturation levels). In view of an extension of this approach towards a non-tenuous beam interacting with a plasma, it is clear that the return current would play an important physical role, not only because its intensity would be large, but also because it would be a source (in a two-dimensional configuration) of a no-longer-negligible magnetic field gradient.

\section{Source of friction and resonance detuning}\label{sec4}

Up to now, we have considered the BPS as unaffected by dissipation phenomena. In what follows, we address a dissipative effect on the beam particles, corresponding to the friction force exerted by the plasma electrons at sufficiently large impact parameter. This phenomenon has the merit of being suitably modelled in the considered scenario, and it can be explicitly evaluated assuming that a cylindrical portion of the plasma, with its axis aligned with the beam trajectory, is removed. The radius of this cylinder is taken larger than the Debye length of the plasma, so that we are excluding short impact parameter collisions in the present study. Such a simplification of the friction term clearly disregards a significant contribution due to local collisions which, however, would unavoidably yield scattering, thus violating the one-dimensionality of the BPS and requiring an extension of the model to more spatial dimensions. Furthermore, we are interested in investigating how friction affects processes like wave-particle resonance and frequency detuning, more than providing a quantitative estimate of the whole friction effect in the plasma.

We observe that, when implementing the BPS in the characterization of the interaction of fast ions with Alfvénic modes, the presence of sources and sinks is often included (see \citers{BB90a,BB90b,BB90c}). Furthermore, such interaction is characterized by a drive, due to the radial gradient of the pressure profile, and a damping rate, due to the background plasma properties. It is just this damping effect that the following analysis is able to mimic, since it also deals with the beam-plasma interaction due to the thermal plasma properties. Thus, in addition to the direct physical interest for the interaction of the beam particles with large-impact-factor thermal electrons, this scenario offers also an improved dynamical picture to be used for predicting features of fast ions in real tokamak devices.

In this respect, to take into account dissipative effects on beam particle dynamics, a new force term is added to the motion equations. In particular, in \citer{NR55}, it was suggested to deal with contributions having a short range and long range separately, and then to consider them in one single force term (there, heavy particles have been considered). Here, for the sake of consistency with the BPS model, we neglect collisions with short range and we focus on the collective effects of the far region of the plasma on the motion of beam particles. As stressed above, such effects are obtained by subtracting a region of the plasma contained in a cylinder of radius $\rho_{min}$ around the trajectory of the electron beam, and then by calculating the potential induced by the injected particles in the far region of plasma, wherein polarization is produced. Then, the field generated by the polarized plasma region on the trajectories is the cause of the friction force undergone by each particle. In this way, in accordance with \citer{NR55}, the force term (specified for the beam electrons) can be expressed as follows:
\begin{align}
F_{fric}=\frac{ e^{2}\rho_{min}}{\omp {x'}^{3}}K_{0} \Big( \frac{\rho_{min}}{{x'}} \Big) K_{1} \Big( \frac{\rho_{min}}{{x'}} \Big)\;,
\end{align}
where $K_{0}$ and $K_{1}$ are the second kind modified Bessel functions of order $0$ and $1$, respectively. The critical parameter $\rho_{min}$ has to obey the following two restrictions:
\begin{align}
p_{\bot} = \frac{2 e^{2}}{\rho_{min}\omp {x'}} \ll m \langle v_e^{2}\rangle^{1/2}\;,
\end{align}
ensuring that the motion of incident beam particles slightly perturbs the plasma unperturbed electrons (having mean square velocity $\langle v_e^{2}\rangle$) at a certain distance $\rho_{min}$ (here $p_\bot$ denotes the momentum gain of the plasma electrons by the passage of the incident particle), and 
\begin{align}
\rho_{min} \gg \lambda_{D}\;,
\end{align}
where $\lambda_D$ denotes the Debye length of the plasma.
\begin{figure}
\centering
\includegraphics[width=\dime]{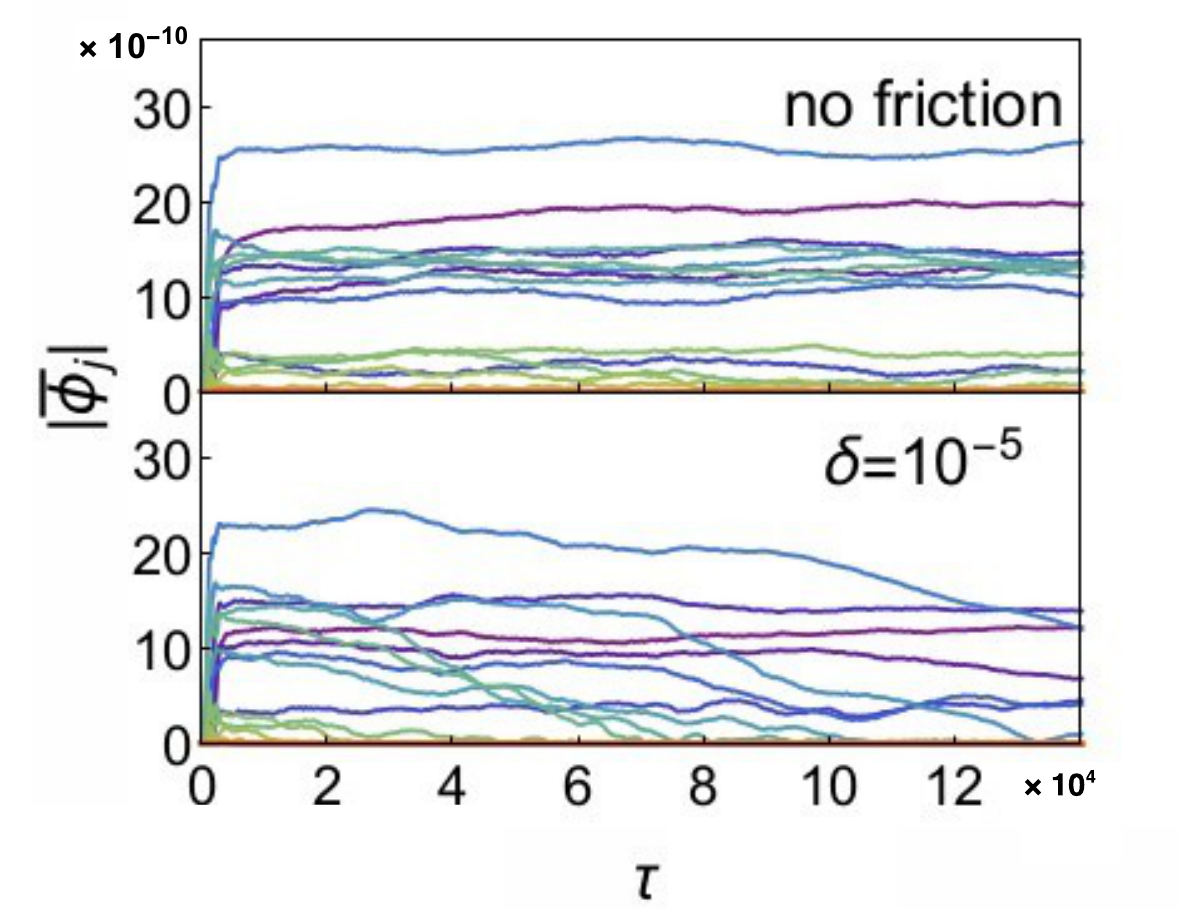}
\caption{Warm beam, multi-mode simulation: mode evolution without dissipation (\eref{mainsys1}) and in the case of a friction coefficient $\delta=10^{-5}$ in \erefs{sgbgdik}, as indicated in the plots. Colour scheme as in \figref{fig-matt-3}. \label{fabio3}}
\end{figure}

Considering a cold plasma, \ie $\omp {x'}\gg\langle v_e^{2}\rangle^{1/2}$, we can assume as reference case $\omp {x'}=100\langle v_e^{2}\rangle^{1/2}$ (thus, the minimum value of $\rho_{min}$ is of order $10\lambda_{D}$). Finally, the set of equations in \reff{mainsys1} can be generalized to include the considered friction effect as follows:
\begin{subequations}\label{sgbgdik}
\begin{align}
\bar{x}_i' &=u_i \;,\\
u'_{i} & = \sum_{j=1}^{M}( \mathrm{i}\ell_{j} \bar{\phi_{j}} \mathrm{e}^{\mathrm{i}\ell_{j}\bar{x}_{i}} + \mathrm{c.c.}) - A \frac{\delta^{4}}{u_{i}^{3} N_{D}} K_{0}\left(B \frac{\delta}{u_{i}}\right) K_{1}\left(B \frac{\delta}{u_{i}}\right),    \\
\bar{\phi}'_{j} &= - \mathrm{i} \bar{\phi}_{j}  + \frac{\mathrm{i} \eta }{2 \ell_{j}^{2} N} \sum_{i=1}^{N} \mathrm{e}^{-\mathrm{i}\ell_{j}\bar{x}_{i} }  \;,
\end{align} 
\end{subequations}
where $A=(2 \upi)^{4}/4 \upi \times 10$, $B=10 \times 2 \upi$ , $\delta=\lambda_{D}/L$ and $N_D$ is the Debye number of the plasma. The parameter $\delta$ weights the relevance of the friction term with respect to the electric force. The range of the values of such a parameter is fixed by the upper bound $\delta \leqslant 10^{-5}$, a consequence of the condition  $u \gg \langle v_e^{2}\rangle^{1/2}$, and by a lower bound due the computational demand of the simulation.
\begin{figure}
\centering
\includegraphics[width=0.95\columnwidth]{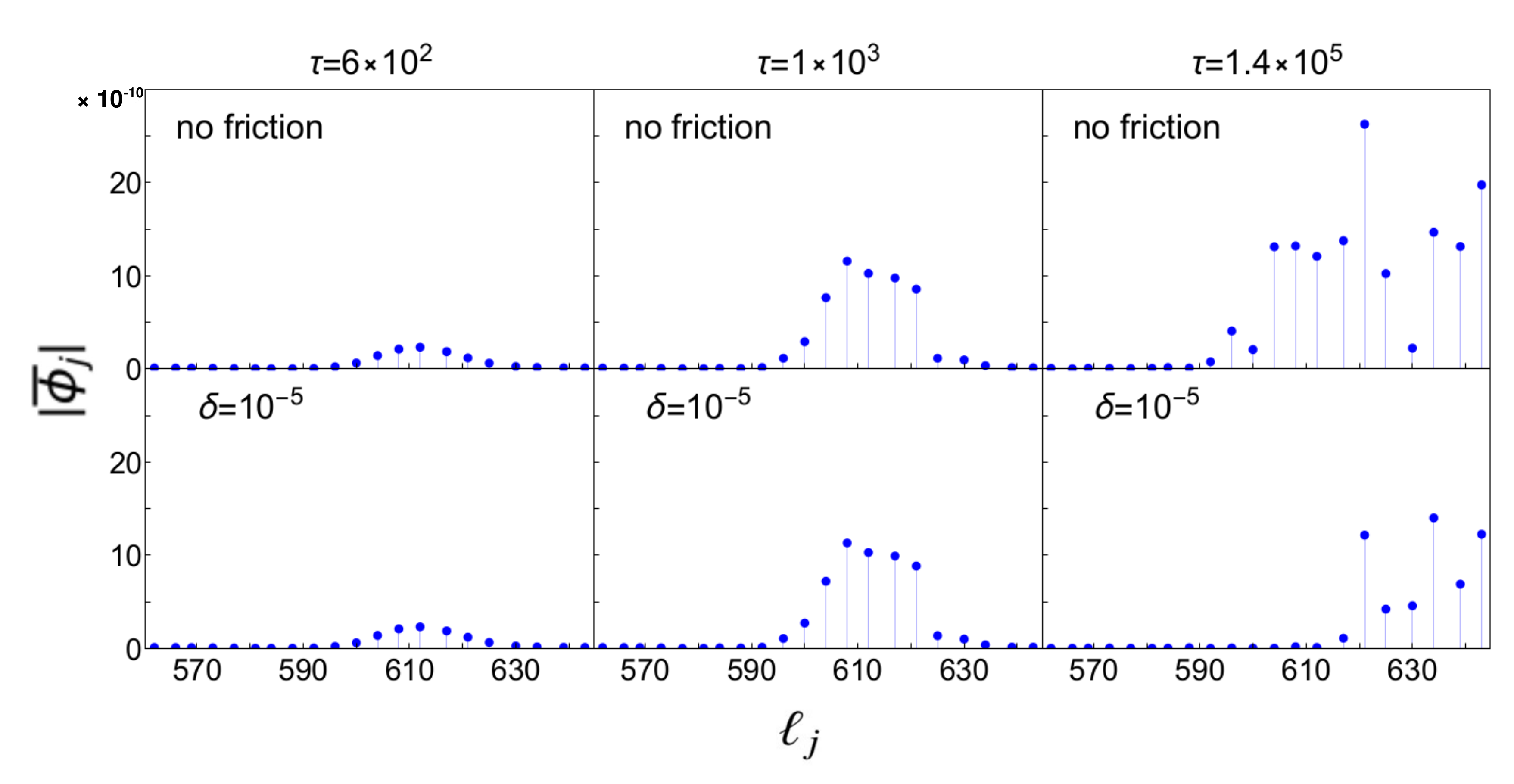}
\caption{Spectral evolution at different times, as indicated above the plots, with and without friction (same as \figref{fabio3}). We note that, as in the previous section, we fixed $\ell_{min}=560$ and $\ell_{max}=640$, resonating in the interval $0.00156\leqslant u \leqslant 0.00178$ (see also \figref{fabio5}).\label{fabio4}}
\end{figure}

\begin{figure}
\centering
\includegraphics[width=0.95\columnwidth]{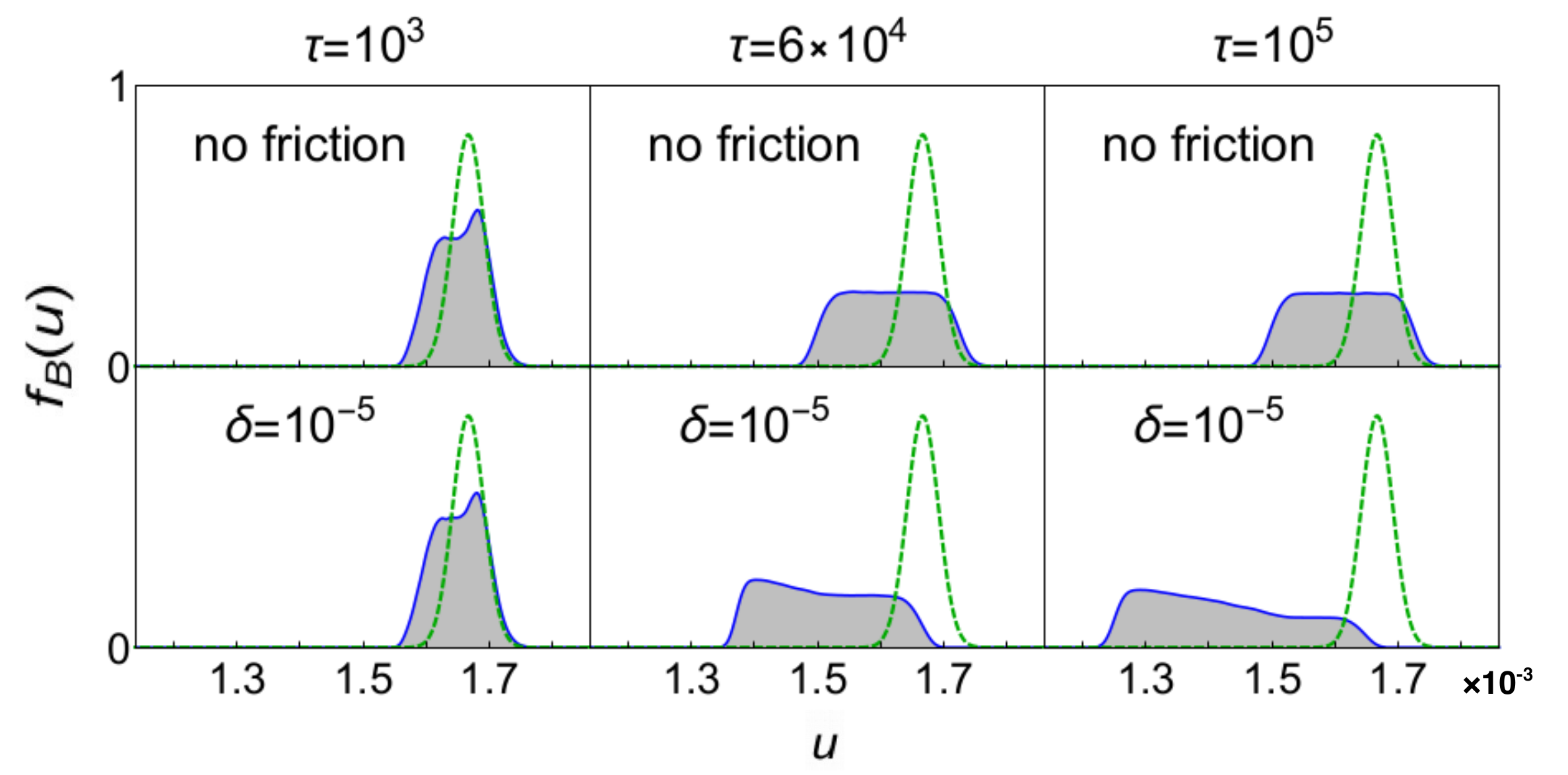}
\caption{Plots of the temporal evolution of the velocity distribution function at different instants, as indicated above the graphs, with and without friction (same as \figref{fabio3}). The green dashed line denotes the initial beam velocity profile. \label{fabio5}}
\end{figure}

In \figref{fabio3}, we plot the spectral evolution for the case of 21 self-interacting modes, with and without the friction term, assuming an initial Gaussian distribution function in the velocity space (as represented in \figref{fabio5}). Other parameters characterizing the simulation are $\eta=7 \times10^{-6}$, $N_{D}=10^{3}$ and again $\ell_{min}\leqslant\ell_j\leqslant\ell_{max}$, having equispaced resonant velocities. It is worth stressing that several modes are suppressed at large time scales, but this is not a general feature since some modes remain unaffected. In order to better understand this feature, we also analyse the behavior of the spectrum, plotted in \figref{fabio4}. First of all, it is notable that, at late times, while the ideal BPS shows a steady state of the spectrum, the considered dissipative model does not reach this limit. 
In fact, Coulomb friction produces a progressive loss of energy in the beam particles, with kinetic energy transferred to the thermal plasma in the form of a very weak re-heating effect (we recall that the beam is here a tenuous one). As a result, the resonance between particles and field is therefore shifted towards lower velocity values, corresponding to higher wavenumbers in the field spectrum (here, the frequency of the plasma is assigned as a fixed parameter) and 
consequently, we observe the deformation of the field spectrum amplitude, as described in \figref{fabio4}.

It is easy to realize that the basic mechanism entering the spectral evolution is a progressive detuning of each resonance due to the particle loss of energy. Clearly, the modes which are resonant with high-velocity particles are the first to be affected by this mechanism along the system evolution. Differently, those modes which exchange power with low-velocity particles are still supported by that amount of them which was interacting with the previous set of modes. The validity of this conjecture is clearly confirmed by \figref{fabio5}, where the distribution function behavior is shown and the dragging of particles in the velocity space is outlined. In fact, the distribution function evolution without friction (upper panels) is characterized by the typical flattening, while, in the presence of the friction term, a new morphology emerges in which a bump corresponding to small velocities is present. 
If the number of available modes at large wavenumbers were infinite, we would see the particle distribution function progressively flatten at low velocity and the kinetic energy of the beam unavoidably dissipated. 
In this sense, a real steady state here cannot exist, but the described effect must be considered as a very slow phenomenon, so that, on the time scale on which the $N$-body approach remains valid (see the discussion in \cite{CFGGMP14} on the so-called quasi-stationary states), 
its effect appears negligible, while still playing a significant role in the resonance phenomenon.

This study is relevant when the BPS model is implemented to mimic the radial transport in a tokamak device. In fact, if EPs lose energy for specific phenomena taking place in the tokamak configuration, this can affect the AE spectrum. For instance, in \citep{spb16}, it is shown that the low toroidal number modes can be destabilized via a cascade process. These modes would be, using the mapping procedure addressed in Sec.\ref{newtok}, those ones surviving in \figref{fabio4}. Thus, including friction term in the BPS model can be important in reproducing specific tokamak features due, for instance, to collisional effects.

\section{Concluding remarks}
We analysed different aspects of the beam-plasma instability clarifying which features of the original treatment have to be improved to adapt this scheme to the interaction between fast ions and the Alfv\'enic spectrum in a tokamak device.

As a first step, we gave a reformulation of the Poisson equation for the BPS to outline the basic hypotheses at the ground of the dielectric assumption which, in the original analysis, was made for the background plasma.

Then, we studied the so-called non-perturbative effects of the beam-plasma dispersion relation, showing how the inverse Landau damping formula fails approaching a flat region of the fast particle distribution function. Actually, in such a condition, a significant frequency shift is naturally predicted by the distribution function and a non-vanishing growth rate emerges from the integral nature of the dispersion relation. 

After that, we analysed the determination of a return current into the bulk of the plasma as an effect of the fast particle crossing. In this respect, we treated the interaction of the beam with the background plasma fluctuations, describing the former still as an $N$-body Hamiltonian system, while the latter in terms of linear fluid dynamics. We then arrived at a restated $N$-body dynamics, whose Poisson equation is now able to account for the interaction of fast particles with the Langmuir waves present in the plasma. We showed how the emergence of a return current implies a suppression of the saturation level of the spectrum and the obtained mode growth rates. This fact suggests that, in the case of a non-tenuous beam, the dielectric approximation, if still adopted, must be amended for effects of the plasma back-reaction that, in turn, influence the beam dynamics and the resulting beam profile.

Finally, we studied the correction to the BPS when the interaction between fast particles and plasma particles far from the beam trajectory is taken into account. A friction term arises due to the plasma charges at scales much greater than the Debye length. This effect determines a significant displacement of the beam particles in the distribution function from high velocities to smaller ones. The resulting field spectrum is initially damped in correspondence to larger wavelengths, and eventually the whole profile is affected.

We stress that, in both these last analyses, the ions of the thermal plasma are considered as essentially at rest, so that they behave here simply as a ``neutralizing background'' for the system. This hypothesis is well grounded for a sufficiently cold plasma, since the ion response takes place on a time scale much slower than both the inverse of electron plasma and bounce frequencies. This scenario is natural for the BPS and easily amendable if the physical conditions require one to account for ion motion. In other words, we are applying to plasma physics the so-called Born-Oppenheimer approximation typical of the physics of matter, when molecular spectra are concerned. In plasma physics, such an assumption lives on a pure classical sector and it is justified by the mass separation existing between ions and electrons. However, it is important to distinguish here between the single-electron velocity, like that of the beam particles and the electron fluid velocity, characterizing the background response via a return current. The single-electron velocity is always much larger that the single-ion one, but the electron fluid, the velocity of which is an average of the single-particle motion, remains in principle small, as happens in our study, due to the validity of quasi-neutrality in the unperturbed plasma at the physical scale of interest.

We conclude that, when implementing the results of this study into real tokamak configurations, the considered effects are not directly relevant: return current and friction are associated with the motion of fast ion beams, but their description requires a different setting with respect to the present one, especially because of the radial non-uniformity of the tokamak equilibrium.

The relevance of the considered deviations from a standard BPS picture lies in the possibility of mimicking corresponding features of the mapped scheme from the velocity space to the radial profile of a tokamak. In particular, while the drive of the instability of fast ions interacting with Alfv\'enic modes is essentially governed by the radial profile of the pressure of the hot particle population, \ie the radial slope of its distribution function, the damping undergone by the ion beam depends on intrinsic features of the background equilibrium plasma. The relevance to dealing with a BPS accounting for damping features relies on modelling this effect when the mapping procedure is implemented.

\vspace{1cm}

{\footnotesize *** This work has been partly carried out within the framework of the EUROfusion Consortium [Enabling Research Projects: NAT (AWP17-ENR-MFE-MPG-01), MET (CfP-AWP19-ENR-01-ENEA-05)] and has received funding from the Euratom research and training programme 2014-2018 and 2019-2020 under grant agreement No 633053. The views and opinions expressed herein do not necessarily reflect those of the European Commission. ***}


\end{document}